\documentclass[a4paper]{PoS}

\usepackage{subfig}   
\title{On the Spectral Shape of Gamma-ray Pulsars Above the Break
   Energy}

\ShortTitle{On the Spectral Shape of Gamma-ray Pulsars Above the Break
   Energy}

\author{\speaker{Christopher Bochenek} \\
        Enrico Fermi Institute, University of Chicago, 5640 S Elllis Ave, Chicago, IL 60637
\\
        E-mail: \email{cbochenek@uchicago.edu}}

\author{Andrew McCann\\
        Enrico Fermi Institute, University of Chicago, 5640 S Elllis Ave, Chicago, IL 60637\\
        Now at The Department of Physics, McGill University, Montreal, QC, H3A 2T8, Canada\\
        E-mail: \email{mccann@hep.physics.mcgill.ca}}

\abstract{
It is well known that, for bright gamma-ray pulsars with high statistics above a few GeV, the phase averaged spectral energy distribution (SED) is harder than a simple exponential cutoff above the break. We perform phase-resolved spectral analyses of bright gamma-ray pulsars and demonstrate that, even over narrow phase ranges, the SEDs of gamma-ray pulsars above the break energy are harder than a simple exponential cutoff. We argue within a radiation-reaction limited curvature framework that this is indicative of non-stationary emission or emission from multiple zones. Further, we address a common problem faced when fitting hard spectral tails with a power-law times a sub-exponential function. Namely, that the sub-exponent parameter does not describe any parameters of physical models of pulsar emission. We introduce a simple analytical fit function to solve this problem.}

\FullConference{The 34th International Cosmic Ray Conference,\\
		30 July- 6 August, 2015\\
		The Hague, The Netherlands}

\begin{document}

\section{Introduction}
The most commonly accepted mechanism to explain the emission from gamma-ray pulsars is curvature radiation, which is radiation from a relativistic charged particle moving along a curved magnetic field line. While the particular details of specific magnetospheric models may vary considerably, it is generally agreed that if the emission originates in the magnetosphere, particles are accelerated in regions where the plasma density is lower than the Goldreich-Julian density. These regions, referred to as gaps or emission zones, can support large electric fields which energise particles. These energetic particles then lose energy via curvature radiation and other processes \cite{Gil, Muslimov, Romani, Vigano}. Magnetospheric models based on curvature radiation can generally reproduce the gamma-ray emission from pulsars observed by Fermi-LAT. Other models, which place the site of emission in the current sheet beyond the light cylinder are also reasonably successful at reproducing the observational data \cite{Petri}. 

Within the curvature radiation framework, the spectrum of the emitted gamma rays follows a power-law with a spectral index related to the energy distribution of the emitting particles \cite{Longair}. Beyond some critical energy, the acceleration gains of an electron are equaled by the radiative losses (the so-called radiation-reaction limit), yielding a gamma-ray spectrum that is exponentially suppressed \cite{Ochelkov}. Thus, for a single stable electron population emitting curvature radiation, the gamma-ray spectrum takes the form of a power-law times an exponential cutoff (PLEC).

Within an outer magnetospheric emission framework, gamma-ray emission beams are expected to by wide due to the large radius of curvature of the magnetic field along which the particles move, and due the fact electrons and positions are accelerated in opposite directions, leading to anti-parallel emission beams. Relativistic effects due to the large co-rotation velocity and the proximity to the neutron star distort the emission beams. These effects, coupled with time-of-fight differences between individual emission sites combine to create phases where emission beams converge (caustics) or diverge. In this way, the geometric framework of outer magnetospheric models directly produce the structure (peaks, bridges, off-regions, etc) seen in pulsar emission profiles. 
%Rather, we expect a sub-exponential cutoff, which is harder and rounder than a simple exponential cutoff and is the superposition of many simple exponential cutoffs [1, 2, 12]. This subexponential shape has been observed in the spectra of several gamma-ray pulsars with high statistics above the break energy [1, 2, 3].

%However, in this work, we show that the spectrum in very narrow phase ranges is better described by a power law with a sub-exponential cutoff (PLSEC) rather than with a power law with a simple exponential cutoff (PLEC) as expected in curvature radiation. Within a radiation-reaction limited curvature radiation framework, this would imply that over narrow phase ranges, there are multiple curvature radiation spectra, and thus a range of spectral indices and break energies. Within this framework, this could indicate the existence of multiple or non-stationary emission zones.
%\par
%Furthermore, when fitting to a sub-exponential cutoff, there is no physical significance to the sub-exponent parameter. Every other fit parameter in this functional form maps to some aspect of the pulsar's emission. We propose to solve this problem with a fit function with four parameters: a normalization, an index, a minimum break energy, and a maximum break energy. This fit function is simply the superposition of many PLECs where each simple exponential cutoff has a different break energy.

Given these considerations, in an outer gap curvature emission model, there is likely no 1-to-1 mapping between the phase of an observed photon and the emission region within the magnetosphere where the photon originated. If this is the case, then one should expect to see evidence for the emission from multiple zones within a phase-resolved
pulsar spectrum. In this work we show, through phase-resolved spectral analyses of bright gamma-ray pulsars, that the spectral shape in a given narrow phase range is more compatible with a power-law times a sub-exponential cutoff (PLSEC) than a PLEC. We argue that this PLSEC shape results from a sum of PLEC spectra and that this is evidence for multi-zone emission. Further, we introduce a simple functional form which is informed by these arguments and show that it can replace the PLSEC form in pulsar spectral fitting.

The remainder of this manuscript is structured in the following way. In Section 2, we explain how we performed our phase averaged and phase resolved spectral analyses. In Section 3, we discuss the results from our analysis and show that over narrow phase ranges, the emission from gamma-ray pulsars is harder than a simple PLEC and better described by a PLSEC. In Section 4, we explain how this is evidence for multizone emission or nonstationary emission zones and introduce a fit function intended to replace the PLSEC in pulsar spectral fitting.
%Explain radiation-reaction limited curvature emission mechanism
%In above explanation, explain an emission zone and why it has a simple exponential cutoff along with physical significance of parameters
%cite Vela paper explaining that index and cutoff vary with pulse phase
%This implies that overall spectrum will be a power law with subexponential cutoff, cite many papers that show this spectral fit and the reasoning that it is the superposition of many simple exponential cutoffs
%Then sumarize what we show and our results
\section{Fermi-LAT Data and Analysis Steps}
%explain in detail exactly what we did
The analysis presented uses 5 years of Pass 7 Reprocessed photon data recorded by the Fermi-LAT. \textit{Source}-class events falling within a 20 degree region-of-interest (ROI) are selected around each pulsar. A zenith angle cut of 100 degees is imposed on all events in order to minimize contamination from gamma-rays created in the Earth's limb.  Phase values are assigned to each photon using the Fermi TEMPO2 plugin and pulsar timing models from Dr.\ Matthew Kerr\footnote{https://confluence.slac.stanford.edu/display/GLAMCOG/LAT+Gamma-ray+Pulsar+Timing+Models} \cite{Ray}. All spectral analyses (phase-averaged, phase-resolved, and energy-band fitting) are performed by applying three iterations of the maximum likelihood fit.  In each step the fit-tolerance condition is increased and sources with a low test statistic at are removed from the fit model. To ensure that final fit model matched the data, residual maps are generated by subtracting a skymap derived from the fit model from a counts map derived from the data. These maps are inspected to ensure the scatter in the residuals is normal and that no large scale spacial structures or anomalies are present.

Phase averaged analyses are performed first, starting with input fit
models derived from the 3FGL catalog\footnote{An additional
  step is performed in the analysis of the Vela pulsar since the
  extended source, Vela-X (3FGL J0833.1-4511e), is spatially
  coincident with the Vela pulsar. The parameters of this source are
  derived by gating the data in the phase-range 0.8-1.0 where there is
  negligible emission from the Vela pulsar, and performing a
  likelihood fit. The model parameters of Vela-X are fixed in all
  subsequent analyses to the value returned from this pulsar-gated
  likelihood analysis, with the necessary scaling applied to the
  prefactor parameter.} \cite{3FGL}. The spectral and normalization parameters of
the pulsar under study are free to float, as are a the normalization
parameters of all sources within 4 degrees of the source. All other
parameters are fixed to the 3FGL value, bar the normalization of the
galactic diffuse model which is also free to float. The best-fit model
from the phase-averaged fit is used as the input model for the
phase-resolved fit. For this analysis all parameters, bar the those
associated with the pulsar, are fixed to the value derived from the
phase-averaged fit\footnote{The prefactor parameters were scaled to
  the size of the phase-gate in question in each fit.}. In each phase
range, the pulsar spectrum is models as both a PLEC function
\begin{equation}
\frac{dF}{dE}=A(E/E_0)^{-\Gamma}e^{-(E/E_c)^b}
\end{equation}
(with the sub/super-exponent $b=1$) and a PLSEC function (with $b$
free). The phase gates for the phase-resolved analyses are chosen to
equalise the number of counts from the pulsar in each phase bin.

To make a spectral energy distribution (SED) from the phase averaged analysis, our data is partitioned into 12 evenly spaced logarithmic bins in energy with 4 bins per decade. On each of these energy bins, the three stage analysis process is applied, modeling each source within 4 degrees of each pulsar as a power law with spectral index fixed to 2. Using the output of this analysis in each energy bin, the energy flux for the pulsar is calculated.
%For each phase averaged analysis, we performed a phase resolved analysis\footnote{The phase resolved analysis was done in 30 phase bins chosen to equalise the number of counts from our source in each bin. Source with fixed model parameters had their prefactors scaled appropritatly.}. For each phase resolved analysis fixed all sources except for the pulsar, and allowed all of the pulsar's spectral parameters to float. We recorded the spectral index, break energy, and, where appropriate, the subexponent value from each phase resolved fit for each phase bin. 
%Furthermore, because the phase averaged spectrum is the sum over all phases of each phase resolved spectrum, we added all of the phase resolved analyses whose subexponent was fixed to unity in order to investigate whether it was sufficient to describe the phase averaged spectrum as a sum over all phases where each phase is described by a PLEC.
\section{Phase Resolved Spectral Shape of Each Pulsar}
\begin{figure}[t]
\centering
%\begin{subfigure}[Geminga]
\includegraphics[width=.9\textwidth]{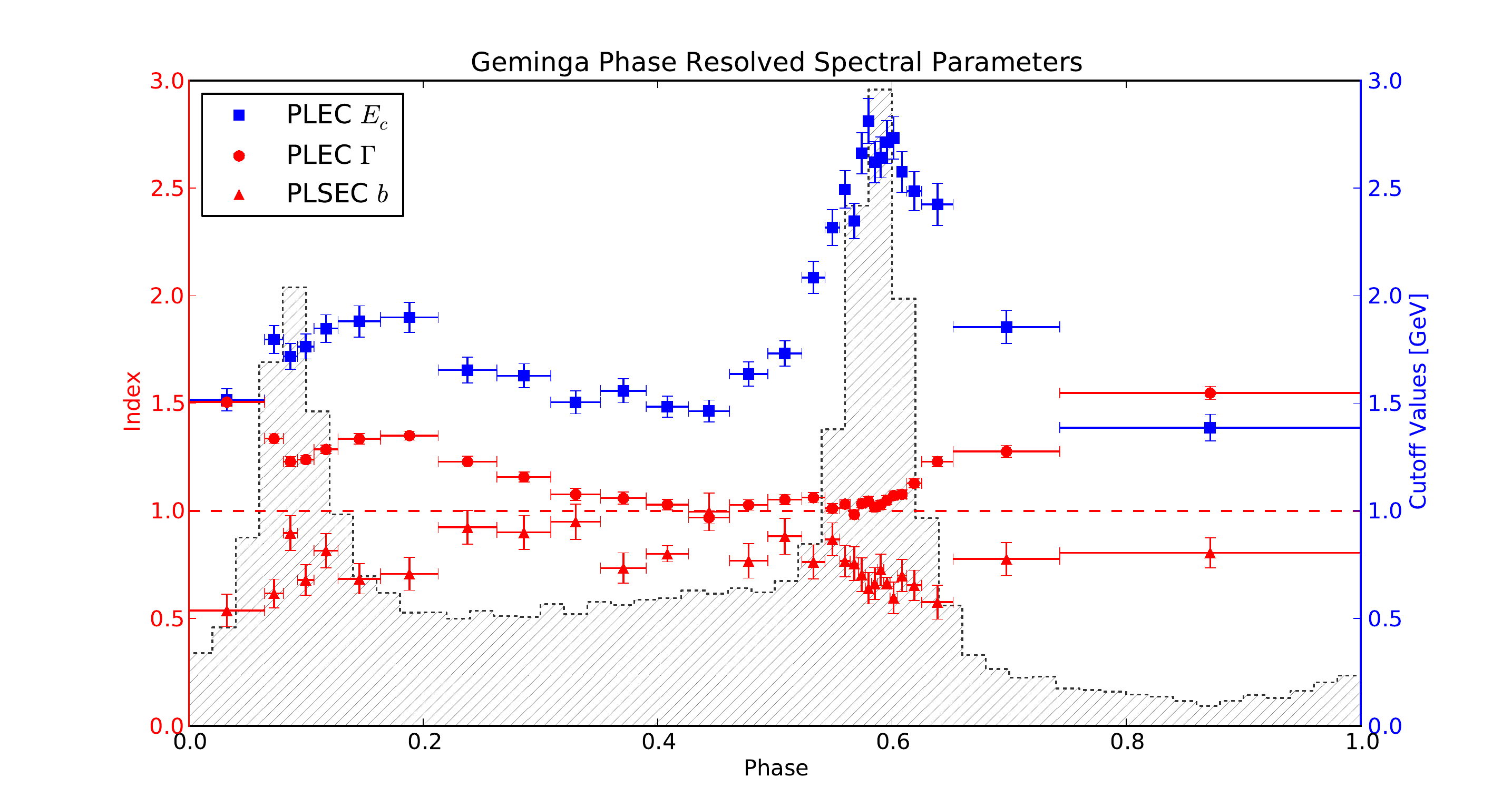}
%\end{subfigure}
\includegraphics[width=.9\textwidth]{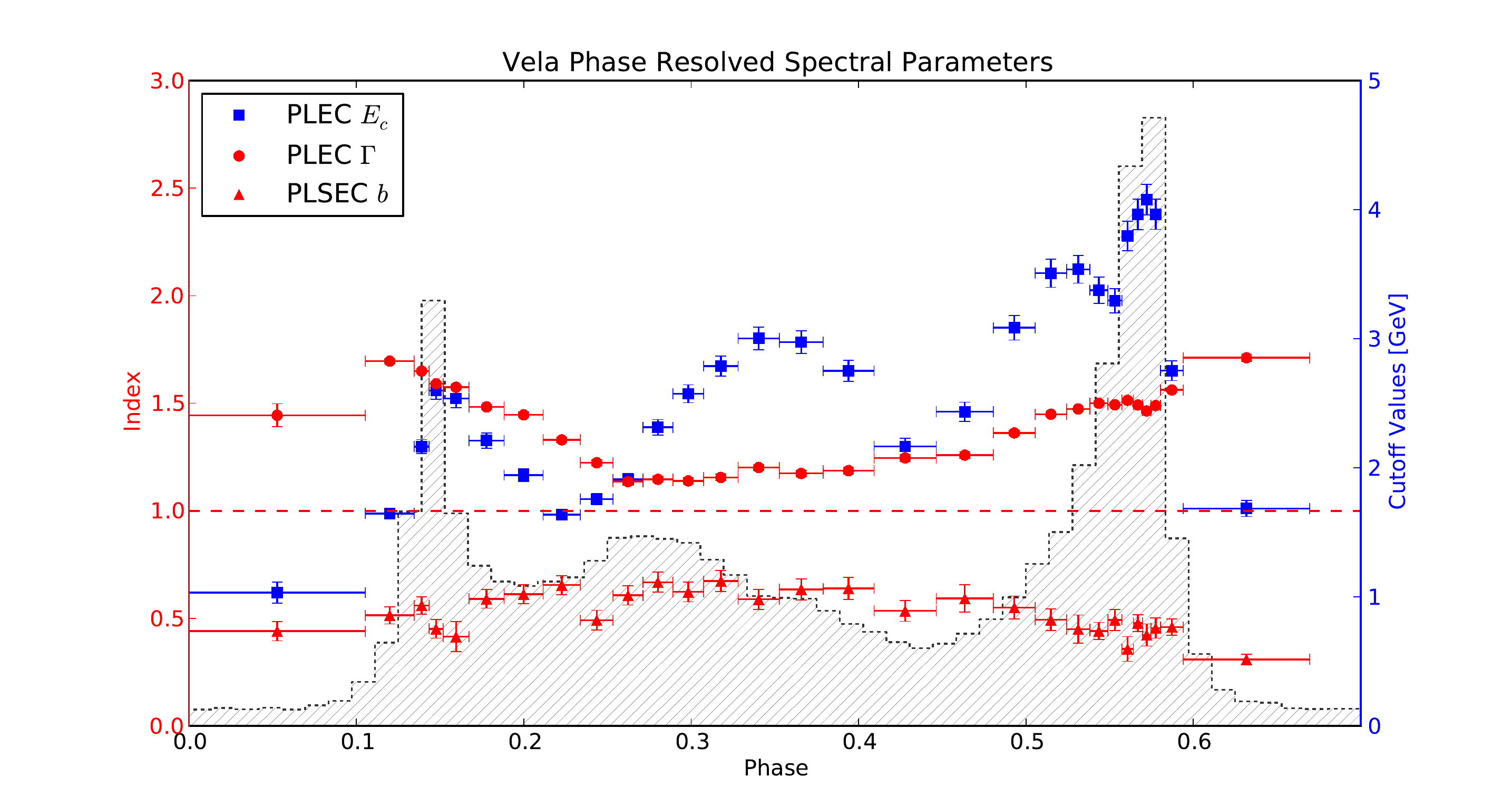}
\caption{The phase-resolved spectral parameters of the Geminga pulsar (top) and Vela pulsar (bottom). The shaded histogram in each figure shows the phasogram of the corresponding pulsar. Note that for Vela, the range on the x-axis is restricted to be between 0 and 0.7. The red circles correspond to the spectral index in the PLEC fit, the blue squares to the PLEC break energy, and the red triangles to the PLSEC subexponent. The red horizontal line indicates an index value of unity.}
\end{figure}
%\begin{figure}[h]
%\centering
%\includegraphics[width=0.75\textwidth]{gemingaSED.png}
%\caption{A phase averaged SED for Geminga. Each black line is the maximum likelihood PLEC fit to a phase bin, while the green line is the sum over each PLEC fit.}
%\end{figure}
Figure  1 plots the phase-resolved spectral parameters of the Geminga and Vela pulsars under both the PLEC and PLSEC model assumptions. In the PLEC fits, the values for the spectral index and break energy in each phase bin are consistent with those obtained from the Fermi collaboration's analysis of both pulsars \cite{geminga,Vela}. From the phase-resolved PLSEC fits to Geminga, we note that the returned $b$ parameter consistently takes a values less than one, and is only compatible with unity in 3 out of the 30 phase bins within the
1-sigma error bar. In the Vela analysis, the $b$ parameter was found to be incompatible with unity in all tested phase gates. We also note that the test statistic values for each pulsar were higher when modeling with a PLSEC than when modeling with a PLEC in each phase range. These results show that the phase-resolved spectra are better described by a PLSEC than by a PLEC. For Geminga, bar the 3 phase regions where $b$ is compatible with unity, we rule out $b=1$ at confidence levels between 1.3$\sigma$ and 12.5$\sigma$, with an average deviation of 3.7$\sigma$. For Vela, we rule out $b=1$ at confidence levels between  6$\sigma$ and 29$\sigma$, with an average deviation of 10$\sigma$.
\par
%However, important part of this plot are the subexponent values. Throughout the entire pulse period, only three of the subexponent values are consistent with a simple exponential cutoff ($b=1$). This indicates that each phase bin is better described by a PLSEC than by a PLEC. This goes against the idea that the phase averaged spectrum has a subexponential tail because it is the sum of curvature radiation spectra in each phase bin, since the curvature radiation spectrum is described by a simple exponential cutoff. 
%\par
%Figure 2 bolsters the argument that the phase averaged spectrum is described by the sum of simple exponential tails in each phase bin. In this plot, the green line is the sum of our phase resolved analyses with $b$ fixed to unity, while the black lines are the spectra of each individual phase bin. The points on this plot are the energy flux values determined from the phase averaged analysis with $b$ fixed to unity. It is clear that the sum of all the phase bins underestimates the phase averaged spectrum at high energies and a harder spectral shape is required. 
%explain our results and show plots and show that they might imply within this regime multiple emission regions or mobile regions
\section{Discussion}Abdo et al.\ \cite{Vela} and Celik \&
Johnson \cite{Celik}
have shown that sub-exponentially suppressed spectra are easily
produced by summing together several pure exponential shapes with a
range of cutoff energies. This, they argue, explains the
sub-exponential cutoff seen in the phase-averaged spectra of some
pulsars, since it is the summation of the phase-resolved spectra which
show significant phase-dependent variation in the cutoff energy (see
Figure~1). We have shown here that, for bright gamma-ray pulsars, the
individual phase-resolved spectra are themselves not exponentially
suppressed, but are better described by sub-exponential breaks. 

Within the context of outer magnetospheric curvature emission models,
we can interpret the appearance of sub-exponential cutoffs in the
phase-resolved spectra as evidence that the emission observed at a
given phase originates from several different particle acceleration
zones, each with a different radiation-reaction energy limit. Given
the wide emission beams and the various relativistic and time-of-fight
effects expected in outer-gap curvature emission, the convergence of
different beams at the same phase is expected and should result in the
type of sub-exponential spectral signature we have observed. Leung et
al.\ \cite{Leung}, have argued that sub-exponential cutoff shapes appear in
pulsars because the accelerating voltage in a given gap is unstable
and thus the time-averaged emission from a single emitting zone is a
summation over various acceleration states. Either interpretation is
compatible with the data. We note that for Geminga, the phases with
the roundest cutoff (where the parameter $b$ takes its smallest value)
occurs in the leading edge of the first peak and the trailing edge of
the second peak while the bridge region, in general, has the sharpest
cutoff ($b$ closest to unity). A similar behavior is seen is Vela. If the emission profiles are interpreted
within an outer gap formalism, the phase-dependent behavior of the $b$
parameter may help shed light on the emission geometry.

%Introduce Andrew's function and explain it in terms of physical parameters
\begin{figure}[t]
\centering
\subfloat[Geminga]{\includegraphics[width=0.5\textwidth]{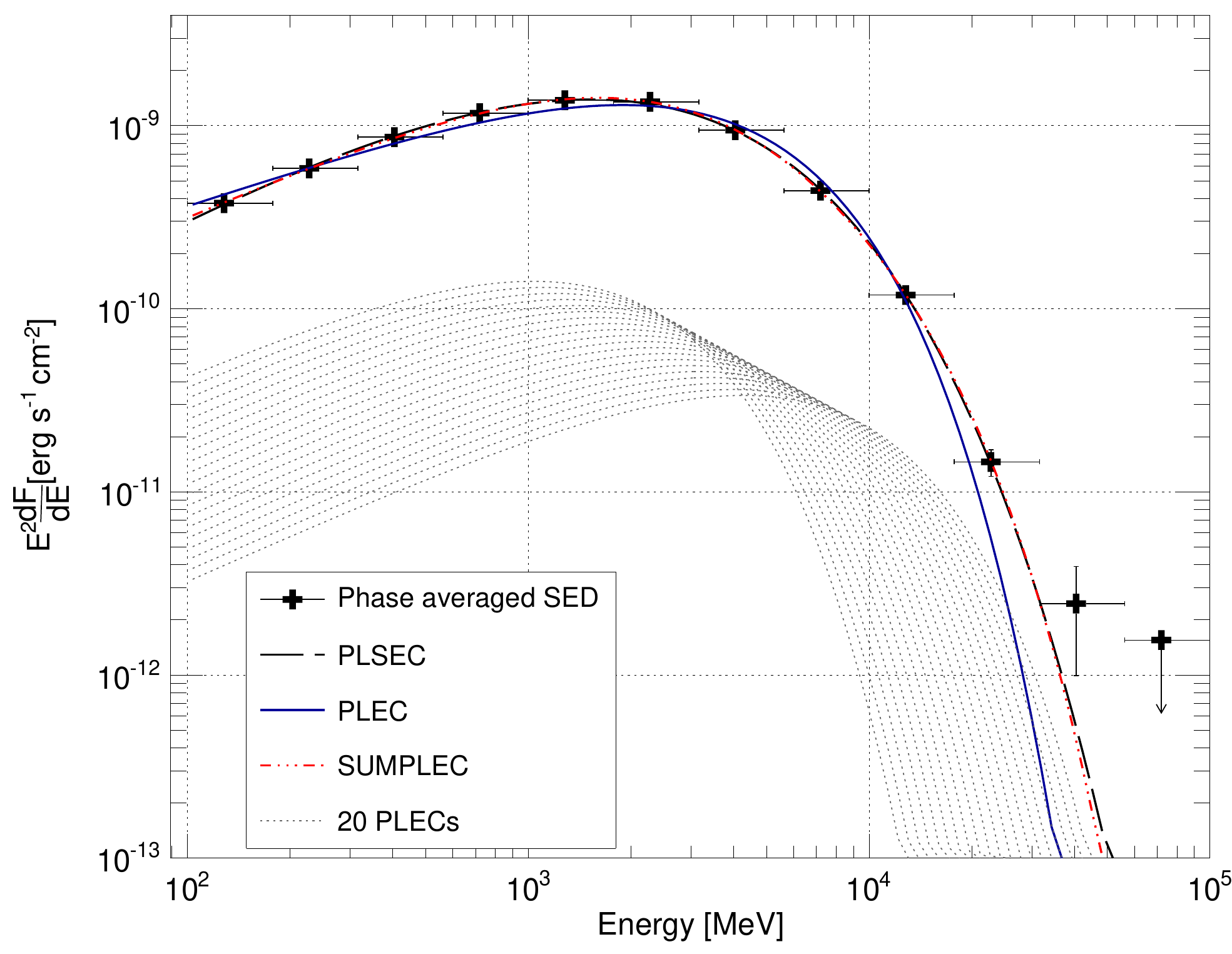}}\hfill
\subfloat[Vela]{\includegraphics[width=0.5\textwidth]{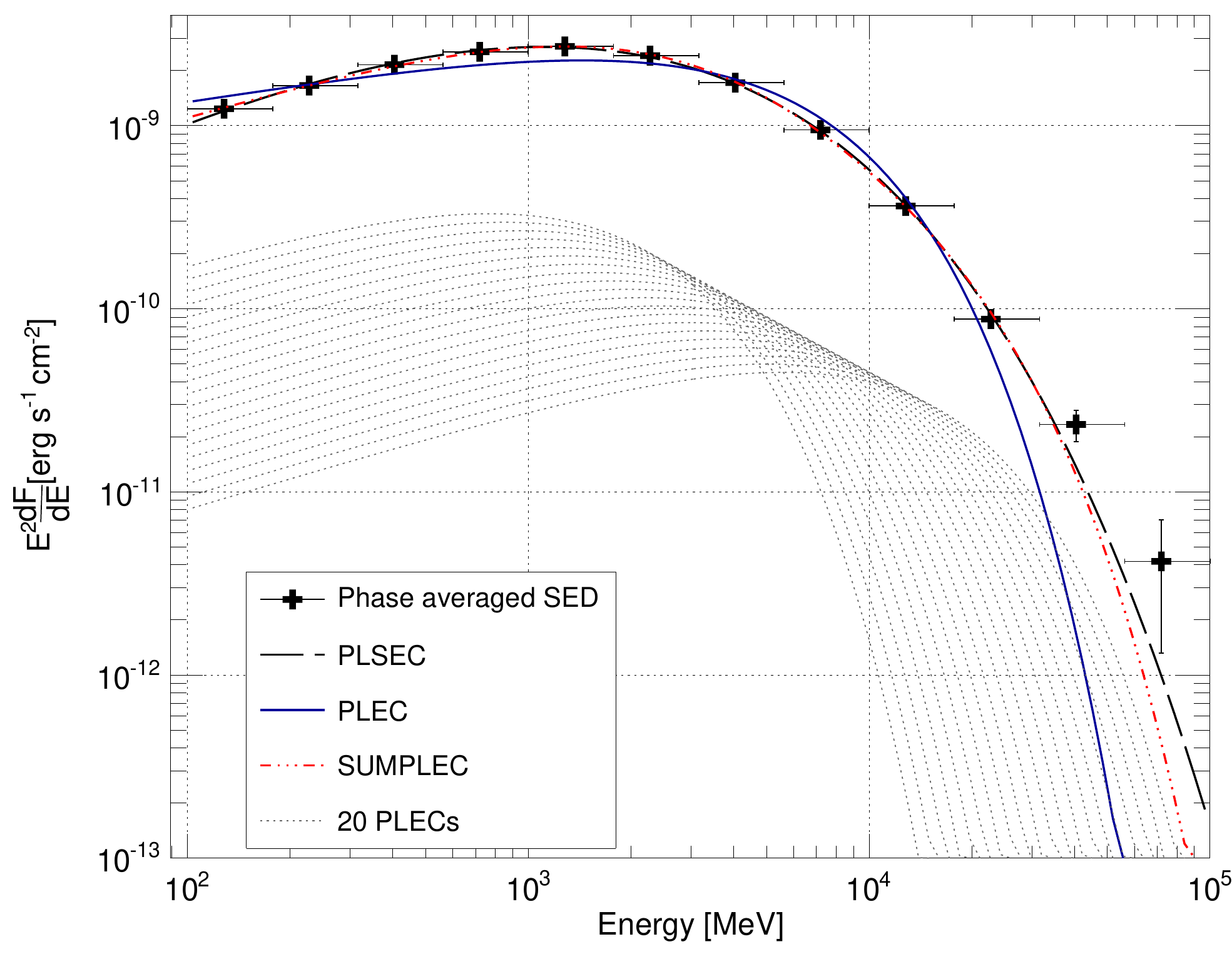}}
\caption{Phase averaged spectral energy distributions of Geminga (left) and Vela (right) that shows the best fit lines from a PLEC, PLSEC, and SUMPLEC.}
\end{figure}
The clear appearance of sub-exponential cutoffs in phase-averaged and,
now, phase-resolved pulsar spectra indicate that the PLSEC functional
form is increasingly required for pulsar spectral fitting. However,
the $b$ parameter in the PLSEC function has no physical analog in
theories high-energy emission and it is highly degenerate with the
cutoff parameter ($E_{c}$). These facts mean that interpreting the
PLSEC fit parameters within typical emission models is impractical. As
discussed, the rounded sub-exponential shape seen in pulsars is
attributed to the sum over several purely exponential cutoffs with a
range of different break energy. With this in mind, we introduce the
following ad-hoc formula:
\begin{equation}
      \frac{dF}{dE} =
      \frac{A}{N}\sum_{i=1}^{N}\left\{\frac{{\left(E/E_0\right)}^{-\Gamma}e^{-\left(E/E^{i}_{\rm
              c}\right)}}{\left(E^{i}_{\rm c}\right)^{1-\Gamma} -
        \left(E_{low}\right)^{1-\Gamma}}\right\}\,\,\,\,,\,\,\,\,\,\,E^{i}_{\rm
        c} = 10^{\left(\alpha + \frac{i}{N}\beta\right)}
\end{equation}
which we call SUMPLEC. This formula is a summation of $N$ PLECs, with
the break energy of each ranging from the value $10^{\alpha}$~MeV to
$10^{\alpha + \beta}$~MeV in $N$ logarithmically spaced steps. The
values of $A$, $\Gamma$, $\alpha$ and $\beta$ are free to
float. $E_{0}$ is the decorrelation energy. The denominator of the
term in the summation has effect of normalizing the area under each
PLEC with the $E_{low}$ parameter fixed to the value of the lower
limit of the fitted energy range. In the case, $N=1$ and $\beta=0$,
this form reduces to the simple PLEC. Using this function to fit the
phase-averaged spectrum of each pulsar, we found that increasing the value
of $N$ asymptotically improved the fit, with the best-fit values
remaining unchanged once $N>5$. To produce Figure 2, we fixed $N$ to
the value 20. Interpreted within an outer gap curvature emission
framework, the values of $10^{\alpha}$~MeV and $10^{\alpha + \beta}$
MeV specify the range of cutoff energies present in the
magnetosphere. For Geminga the cutoff energies span the range $1.22 \pm 0.11$ to $5.1\pm 0.2$ GeV. For Vela, the cutoff energies span the range $1.35\pm 0.13$ to $9.8 \pm 0.5$ GeV.

Figure 2 clearly shows that the SUMPLEC function reproduces a shape very similar to the PLSEC and that it can be used to fit pulsar spectra. By construction,  the parameters of the SUMPLEC function are easily interpreted with outer magnetospheric emission models and, thus, we argue it can replace the  PLSEC function in pulsar spectral fitting.

%\begin{figure}[t]
%\centering
%\includegraphics[width=0.75\textwidth]{my_geminga_sed_only.pdf}
%\caption{A spectral energy distribution of Geminga that shows the best fit lines from a PLEC, PLSEC, and SUMPLEC.}
%\end{figure}
%Introduce Andrew's function and explain it in terms of physical parameters

\end{document}